\documentclass[aps,prd,preprintnumbers]{revtex4}
\usepackage{graphicx}                              
\usepackage{amsmath,amsfonts}
\graphicspath{{./pictures/}}                       
 
\usepackage{dsfont}   
\usepackage{epsfig}
\usepackage{amssymb}
\usepackage{amsfonts}
\usepackage{float}
\usepackage{bbm}
\usepackage{psfrag}
\newcommand{\vs}{\vspace}
\newcommand{\hs}{\hspace}

\newcommand{\bdm}{\begin{displaymath}}
\newcommand{\edm}{\end{displaymath}}
\newcommand{\beq}{\begin{equation}}
\newcommand{\eeq}{\end{equation}}
\newcommand{\bea}{\begin{eqnarray}}
\newcommand{\eea}{\end{eqnarray}}
\newcommand{\bit}{\begin{itemize}}
\newcommand{\eit}{\end{itemize}}
\newcommand{\bc}{\begin{center}}
\newcommand{\ec}{\end{center}}
\newcommand{\re}{\relax{\rm I\kern-.18em R}}
\newcommand{\ID}{\mathbbm{1}}

\newcommand{\fhs}[1]{\mbox{\hs{#1}}}
\newcommand{\ie}{{\it i.e. }}
\newcommand{\Dov}{{\cal D}^{(ov)}}
\newcommand{\D}{\Dov}

\newcommand{\SD}{{\cal D}^{(ov)}}

\newcommand{\Dw}{{\cal D}^{(W)}}
\newcommand{\SDw}{{\cal D}^{(W)}}

\newcommand{\sumFL}{\sum\limits_{i=1}^{N_c}}

\newcommand{\fermiMat}{{\cal M}}

\newcommand{\latVol}[2]{\ifnum #1=#2 $#1^4$ \else $#1^3\times #2$\fi}
\newcommand{\lattice}[2]{\ifnum #1=#2 $#1^4$-lattice \else $#1^3\times #2$-lattice\fi}
\newcommand{\latticeX}[3]{\ifnum #1=#2 $#1^4$-lattice#3 \else $#1^3\times #2$-lattice#3\fi}
\newcommand{\lattices}[2]{\ifnum #1=#2 $#1^4$-lattice \else $#1^3\times #2$-lattices\fi}
\newcommand{\eq}[1]{Eq.~(\ref{#1})}
\newcommand{\eqs}[2]{Eqs.~(\ref{#1}-\ref{#2})}
\newcommand{\sect}[1]{section~\ref{#1}}

\newcommand{\fig}[1]{Fig.~\ref{#1}}
\newcommand{\tab}[1]{Tab.~\ref{#1}}
\newcommand{\Ref}[1]{Ref.~\cite{#1}}
\newcommand{\Refs}[1]{Refs.~\cite{#1}}
\newcommand{\GEV}[1]{#1\,\mbox{GeV}}
\newcommand{\TEV}[1]{#1\,\mbox{TeV}}

\newcommand{\Nconf}{N_{Conf}}

\newcommand{\proz}[1]{#1\,\%}
\newcommand{\Tr}{\mbox{Tr}}
\newcommand{\RE}{\mbox{Re}}

\newcommand{\SUtwoTimesUoneY}{{SU(2)_\mathrm{L}\times U(1)_\mathrm{Y}}}

\newcommand{\dslash}{\ensuremath\partial\kern-0.53em/}

\newcommand{\includeFigDouble}[7]{
\vs{-#6mm}
\bc
\begin{figure}[htb]
\centering
\begin{tabular}{cc}
\includegraphics[width=0.48\textwidth]{#1}
&
\includegraphics[width=0.48\textwidth]{#2}
\\
\hs{4mm}(a) & \hs{8mm}(b)  \\
\end{tabular}
\caption[#5]{#4}
\label{#3}
\vs{-2mm}
\end{figure}
\ec
\vs{-6mm}
\vs{-#7mm}
}

\newcommand{\includeTab}[5]{
\begin{table}[htb]
\centering
\begin{tabular}{#1}
\hline
#2
\hline
\end{tabular}
\caption[#5]{#4}
\label{#3}
\end{table}
}

\newcommand{\includeFigTripleDouble}[9]{
\bc
\begin{figure}[htb]
\centering
\begin{tabular}{ccc}
\includegraphics[width=0.32\textwidth]{#1}\hs{-0mm}
&
\includegraphics[width=0.32\textwidth]{#2}\hs{-0mm}
&
\includegraphics[width=0.32\textwidth]{#3}\hs{-0mm}
\\
\hs{9mm}(a) & \hs{9mm}(b) & \hs{9mm}(c) \\
\hs{9mm}    & \hs{9mm}    & \hs{9mm}    \\
\includegraphics[width=0.32\textwidth]{#4}\hs{-0mm}
&
\includegraphics[width=0.32\textwidth]{#5}\hs{-0mm}
&
\includegraphics[width=0.32\textwidth]{#6}\hs{-0mm}
\\
\hs{9mm}(d) & \hs{9mm}(e)  & \hs{9mm}(f) \\
\end{tabular}
\caption[#9]{#8}
\label{#7}
\vs{-2mm}
\end{figure}
\ec
\vs{-6mm}
}

\begin{document}
\preprint{HU-EP-10/66, DESY 10-180}

\title{Higgs boson mass bounds in the presence of a very heavy\\fourth quark generation} 

\author{P. Gerhold$^{a,b}$, K. Jansen$^b$, J. Kallarackal$^{a,b}$}
\affiliation{$^a$Humboldt-Universit\"at zu Berlin, Institut f\"ur Physik, 
Newtonstr. 15, D-12489 Berlin, Germany\\
$^b$NIC, DESY,\\
 Platanenallee 6, D-15738 Zeuthen, Germany}

\date{November 7, 2010}

\begin{abstract}
We study the effect of a potential fourth quark generation on the upper and lower Higgs boson mass bounds. 
This investigation is based on the numerical evaluation of a chirally invariant lattice Higgs-Yukawa model emulating 
the same Higgs-fermion coupling structure as in the Higgs sector of the electroweak Standard Model. In particular, 
the considered model obeys a Ginsparg-Wilson version of the underlying $\mbox{SU}(2)_L\times \mbox{U}(1)_Y$ symmetry, 
being a global symmetry here due to the neglection of gauge fields in this model. We present our results 
on the modification of the upper and lower Higgs boson mass bounds induced by the presence of a hypothetical very heavy 
fourth quark doublet. Finally, we compare these findings to the standard scenario of three fermion generations.
\end{abstract}


\maketitle

\section{Introduction}
\label{sec:Introduction}
 
The Sakharov explanation for the matter anti-matter asymmetry of the universe suffers from the CP-violating phase 
of the Standard Model (SM3) falling short by at least 10 orders of magnitude. In addition to this concern the Sakharov 
picture demands a sufficiently strong first order electroweak phase transition, which is also objected in the framework 
of the SM3. However, both of the above caveats might be addressable~\cite{Holdom:2009rf,Carena:2004ha} by the inclusion 
of a new fourth fermion generation into an extended version of the Standard Model (SM4). Despite the arguments against 
the existence of a fourth fermion generation such a scenario nevertheless remains attractive for two reasons. Firstly, 
there is a strong conceptual interest, since a new fermion generation would need to be very heavy, leading to rather large 
Yukawa coupling constants and thus to potentially strong interactions with the scalar sector of the theory. Secondly,
it has been argued~\cite{Holdom:2009rf,Eberhardt:2010bm} (and the references therein) that the fourth fermion generation 
is actually {\it not} excluded by electroweak precision measurements, thus leaving the potential existence of a new 
fermion generation an open question.

In our contribution, however, we do not present any statement arguing in favour or disfavour of a new fermion generation. Here, 
we simply assume its existence and focus on the arising consequences on the Higgs boson mass spectrum. With the advent
of the LHC this question will become of great phenomenological interest, since the Higgs boson mass bounds, in particular
the lower bound, depend significantly on the heaviest fermion mass. Demonstrating this effect will be the main
objective of the present work. 

Due to the large Yukawa coupling constants of the fourth fermion generation a non-perturbative computation is highly 
desirable. For this purpose we employ a lattice approach to investigate the strong Higgs-fermion interaction.
In fact, we follow here the same lattice strategy that has already been used in \Ref{Gerhold:2009ub,Gerhold:2010bh,Gerhold:THESIS} 
for the non-perturbative determination of the upper and lower Higgs boson mass bounds in the SM3. This latter approach
has the great advantage over the preceding lattice studies of Higgs-Yukawa models (see eg. 
\Refs{Smit:1989tz,Shigemitsu:1991tc,Golterman:1990nx,book:Jersak,book:Montvay,Golterman:1992ye,Jansen:1994ym}
and the references therein) that it is the first being based on 
a consistent formulation of an exact lattice chiral symmetry~\cite{Luscher:1998pq}, which allows to emulate the chiral 
character of the Higgs-fermion coupling structure of the Standard Model on the lattice
in a conceptually fully controlled manner. The interest in lattice Higgs-Yukawa models has therefore recently been
renewed~\cite{Bhattacharya:2006dc,Giedt:2007qg,Poppitz:2007tu,Gerhold:2007yb,Gerhold:2007gx,Fodor:2007fn,Gerhold:2009ub,Gerhold:THESIS,Gerhold:2010bh}.

The actual details of the considered lattice model will be given in \sect{sec:model}. In the following \sect{sec:simstrag}
the paper elaborates on the pursued simulation strategy for the determination of the upper and lower Higgs boson mass bounds, 
\ie the selection procedure for the bare lattice parameters. The obtained lattice results together with their extrapolations 
to the infinite volume limit are then presented and discussed in \sect{sec:numericalresults} before the paper ends with some
outlook and conclusions in \sect{sec:conclusions}.

\section{The Model}
\label{sec:model}

In order to evaluate the Higgs boson mass bounds we have implemented a lattice 
model of the pure Higgs-fermion sector of the Standard Model. To be more precise,
the Lagrangian of the targeted Euclidean continuum model we have in mind is given as
\begin{align}
\label{eq:StandardModelYuakwaCouplingStructure}
L_{HY} &= \bar t' \dslash t' + \bar b' \dslash b' + 
\frac{1}{2}\partial_\mu\varphi^{\dagger} \partial_\mu\varphi
+ \frac{1}{2}m_0^2\varphi^{\dagger}\varphi + \lambda\left(\varphi^{\dagger}\varphi\right)^2 
+ y_{b'} \left(\bar t', \bar b' \right)_L \varphi b'_R + y_{t'} \left(\bar t', \bar b' \right)_L \tilde\varphi t'_R  \notag\\
& + \mbox{c.c. of Yukawa interactions,} 
\end{align}
where we have constrained ourselves to the consideration of the heaviest quark
doublet, \ie the fourth generation doublet, which is labeled here $(t',b')$.
This restriction is reasonable, since the dynamics of the complex scalar doublet 
$\varphi$ ($\tilde \varphi = i\tau_2\varphi^*,\, \tau_i:\, \mbox{Pauli-matrices}$)
is dominated by the coupling to the heaviest fermions. For the same reason we also 
neglect any gauge fields in this approach. The quark fields nevertheless have
a colour index which actually leads to $N_c=3$ identical copies of the fermion doublet 
appearing in the model. However, for a first exploratory study of the fermionic influence
on the Higgs boson mass bounds we have set $N_c$ to 1 for simplicity.

Since the Yukawa interaction has a chiral structure, it is important to establish chiral symmetry also in 
the lattice approach. This has been a long-standing obstacle, which was finally found to be circumventable by 
constructing the lattice equivalent of $\dslash$ as well as the left- and right-handed components of the quark fields 
$t'_{L,R}$, $b'_{L,R}$ on the basis of the Neuberger overlap operator~\cite{Luscher:1998pq, Neuberger:1997fp,Neuberger:1998wv}. 
Following the proposition in \Ref{Luscher:1998pq} we have constructed a lattice Higgs-Yukawa model with a global 
$SU(2)_L \times U(1)_Y$ symmetry.

The fields considered in this model are the aforementioned doublet $\varphi$ as well as 
$N_c$ quark doublets represented by eight-component spinors $\bar\psi^{(i)}\equiv (\bar t'^{(i)}, \bar b'^{(i)})$, 
$i=1,...,N_c$. With $\D$ denoting the Neuberger overlap operator the fermionic action $S_F$ can
be written as
\bea
\label{eq:DefYukawaCouplingTerm}
S_F = \sumFL\,
\bar\psi^{(i)}\, \fermiMat\, \psi^{(i)}, 
&\quad&
\fermiMat = \D + 
P_+ \phi^\dagger \fhs{1mm}\mbox{diag}\left(y_{t'}, y_{b'}\right) \hat P_+
+ P_- \fhs{1mm}\mbox{diag}\left(y_{t'}, y_{b'}\right) \phi \hat P_-,
\eea
where $y_{t',b'}$ denote the Yukawa coupling constants and the scalar field $\varphi_x$ has been rewritten here 
as a quaternionic, $2 \times 2$ matrix $\phi^\dagger_x = (\tilde \varphi_x, \varphi_x)$, with $x$ denoting the 
site index of the $L_s^3\times L_t$-lattice. The left- and right-handed projection operators $P_{\pm}$ and the 
modified projectors $\hat P_{\pm}$ are given as
\bea
P_\pm = \frac{1 \pm \gamma_5}{2}, \quad &
\hat P_\pm = \frac{1 \pm \hat \gamma_5}{2}, \quad &
\hat\gamma_5 = \gamma_5 \left(\ID - \frac{1}{\rho} \D \right).
\eea
The free Neuberger overlap operator can be written as
\bea
\label{eq:DefOfNeuberDiracOp}
\SD &=& \rho \left\{1+\frac{ A}{\sqrt{ A^\dagger  A}}   \right\},
\quad A = \SDw - \rho, \quad 0 < \rho < 2r
\eea
where $\rho$ is a free, dimensionless parameter within the specified constraints that determines the radius of the circle
formed by the eigenvalues of $\D$ in the complex plane. The operator $\SDw$ denotes here the Wilson Dirac 
operator defined as 
\beq
\label{eq:DefOfWilsonOperator}
\Dw = \sum\limits_\mu \gamma_\mu \nabla^s_\mu - \frac{r}{2} \nabla^b_\mu\nabla^f_\mu,
\eeq
where $\nabla^{f,b,s}_\mu$ are the forward, backward and symmetrized lattice nearest neighbour difference operators in 
direction $\mu$, while the so-called Wilson parameter $r$ is chosen here to be $r=1$ as usual.

The overlap operator was proven to be local in a field theoretical sense also in the presence of QCD gauge fields
at least if the latter fields obey certain smoothness conditions~\cite{Hernandez:1998et,Neuberger:1999pz}. The locality properties 
were found to depend on the parameter $\rho$ and the strength of the gauge coupling constant. 
At vanishing gauge coupling the most local operator was shown to be obtained at $\rho=1$. Here, the notion 'most local' 
has to be understood in the sense of the most rapid exponential decrease with the distance $|x-y|$ of the coupling strength 
induced by the matrix elements $\SD_{x,y}$ between the field variables at two remote space-time points $x$ and $y$. For 
that reason the setting $\rho=1$ will be adopted for the rest of this work.

The introduced action now obeys an exact global $\mbox{SU}(2)_L\times \mbox{U}(1)_Y$ 
lattice chiral symmetry. For $\Omega_L\in \mbox{SU}(2)$ and $\epsilon\in \re$ the action is invariant under the transformation
\bea
\label{eq:ChiralSymmetryTrafo1}
\psi\rightarrow  U_Y \hat P_+ \psi + U_Y\Omega_L \hat P_- \psi,
&\quad&
\bar\psi\rightarrow  \bar\psi P_+ \Omega_L^\dagger U_{Y}^\dagger + \bar\psi P_- U^\dagger_{Y}, \\
\label{eq:ChiralSymmetryTrafo2}
\phi \rightarrow  U_Y  \phi \Omega_L^\dagger,
&\quad&
\phi^\dagger \rightarrow \Omega_L \phi^\dagger U_Y^\dagger
\eea
with the compact notation $U_{Y} \equiv \exp(i\epsilon Y)$ denoting the representations of the 
global hypercharge symmetry group $U(1)_Y$ for the respective field it is acting on. In the continuum 
limit \eqs{eq:ChiralSymmetryTrafo1}{eq:ChiralSymmetryTrafo2} eventually recover the (here global) 
continuum $\mbox{SU}(2)_L\times \mbox{U}(1)_Y$ chiral symmetry.

Finally, the purely bosonic part $S_\varphi$ of the total lattice action $S=S_F+S_\varphi$ is given by the usual 
lattice $\varphi^4$-action
\bea
\label{eq:ContinuumPhiAction}
S_\varphi &=& \sum_{x} \frac{1}{2}\nabla^f_\mu\varphi_x^{\dagger} \nabla^f_\mu\varphi_x
+ \frac{1}{2}m_0^2\varphi_x^{\dagger}\varphi_x + \lambda\left(\varphi_x^{\dagger}\varphi_x\right)^2   ,
\eea
with the bare mass $m_0$, the forward difference operator $\nabla^f_\mu$ in direction $\mu$, and the bare quartic 
coupling constant $\lambda$. For the practical lattice implementation, however, a reformulation of \eq{eq:ContinuumPhiAction}
in terms of the hopping parameter $\kappa$ and the lattice quartic coupling constant $\hat\lambda$ proves to be more
convenient. It reads
\beq
\label{eq:LatticePhiAction}
S_\Phi = -\kappa\sum_{x,\mu} \Phi_x^{\dagger} \left[\Phi_{x+\mu} + \Phi_{x-\mu}\right]
+ \sum_{x} \Phi^{\dagger}_x\Phi_x + \hat\lambda \sum_{x} \left(\Phi^{\dagger}_x\Phi_x - N_c \right)^2,
\eeq
and is fully equivalent to \eq{eq:ContinuumPhiAction}. This alternative formulation opens the possibility of explicitly
studying the limit $\lambda = \infty$ on the lattice. The aforementioned connection can be established through a rescaling 
of the scalar field $\Phi_x \in \re^4$ and the involved coupling constants according to
\beq
\label{eq:RelationBetweenHiggsActions}
\varphi_x = \sqrt{2\kappa}
\left(
\begin{array}{*{1}{c}}
\Phi_x^2 + i\Phi_x^1\\
\Phi_x^0-i\Phi_x^3\\ 
\end{array}
\right), 
\quad
\lambda=\frac{\hat\lambda}{4\kappa^2}, \quad
m_0^2 = \frac{1 - 2N_c\hat\lambda-8\kappa}{\kappa}.
\eeq

\section{Simulation Strategy}
\label{sec:simstrag}

Due to the triviality of the Higgs sector the targeted Higgs boson mass bounds actually depend on the 
non-removable, intrinsic cutoff parameter $\Lambda$ of the considered Higgs-Yukawa theory, which can be
defined as the inverse lattice spacing, \ie $\Lambda=1/a$. To determine these cutoff dependent bounds
for a given phenomenological scenario, \ie for given hypothetical masses of the fourth quark generation,
the strategy is to evaluate the maximal interval of Higgs boson masses attainable within the framework of 
the considered Higgs-Yukawa model being in consistency with this phenomenological setup. The free parameters 
of the model, being the bare scalar mass $m_0$, the bare quartic coupling constant $\lambda$, and the bare 
Yukawa coupling constants $y_{t',b'}$ thus have to be tuned accordingly. The idea for the latter fixation 
is to employ the assumed fourth generation quark masses $m_{t',b'}$ as well as the phenomenological 
knowledge of the renormalized vacuum expectation value of the scalar field $\varphi$ (vev). The latter is
used to determine the physical scale $1/a$ according to
\bea
\label{eq:FixationOfPhysScale}
246\, \mbox{GeV} &=& \frac{v_r}{a} \equiv \frac{v}{\sqrt{Z_G}\cdot a},
\eea
where $Z_G$ denotes the Goldstone renormalization constant and $v$ is the lattice vev.

Concerning the hypothetical masses of the fourth fermion generation quarks, we target here a mass degenerate setup 
with $m_{t'}/a=m_{b'}/a=\GEV{700}$, which is somewhat above its tree-level unitarity upper bound~\cite{Chanowitz:1978mv}.
The degenerate setting, being unproblematic from a numerical perspective, is owed here to the existence of a fluctuating 
complex phase in the opposite scenario, \ie in the non-degenerate case~\cite{Gerhold:THESIS}. However, it is remarked that 
we are currently also investigating a set of other mass settings to study in particular the quark mass dependence of the 
Higgs boson mass bounds.

In lack of an additional matching condition a one-dimensional freedom is left open here, which can be parametrized
in terms of the quartic coupling constant $\lambda$. This freedom finally leads to the emergence of upper and lower
bounds on the Higgs boson mass. As expected from perturbation theory, one also finds numerically~\cite{Gerhold:2010bh} 
that the lightest and heaviest Higgs boson masses are obtained at vanishing and infinite bare quartic coupling constant, 
respectively. The lower mass bound will therefore be obtained at $\lambda=0$, while $\lambda=\infty$ will be adjusted to 
derive the upper bound.

However, in the given lattice model the expectation value $\langle \varphi \rangle$ would always be identical to zero due to the 
symmetries in \eqs{eq:ChiralSymmetryTrafo1}{eq:ChiralSymmetryTrafo2}. The identification $v\equiv \langle \varphi \rangle$
would therefore not yield meaningful results in \eq{eq:FixationOfPhysScale}. The reason is that the lattice averages over 
{\it all} ground states of the theory, not only over that one which Nature has selected in the broken phase. To study 
the mechanism of spontaneous symmetry breaking nevertheless, one usually introduces an external current $J$, selecting then
only one particular ground state. This current is finally removed after taking the thermodynamic limit, leading then to the 
existence of symmetric and broken phases with respect to the order parameter $\langle \varphi\rangle$ as desired.
An alternative approach, which was shown to be equivalent in the thermodynamic 
limit~\cite{Hasenfratz:1989ux,Hasenfratz:1990fu,Gockeler:1991ty}, is to define the vacuum expectation value $v$ as the 
expectation value of the {\textit{rotated}} field $\varphi^{rot}$ given by a global transformation of the original field $\varphi$ 
according to
\beq
\label{eq:GaugeRotation}
\varphi^{rot}_x = U[\varphi] \varphi_x
\eeq
with the $\mbox{SU}(2)$ matrix $U[\varphi]$ selected for each configuration of field variables $\{\varphi_x\}$
such that 
\beq
\label{eq:GaugeRotationRequirement}
\sum\limits_x \varphi_x^{rot} = 
\left(
\begin{array}{*{1}{c}}
0\\
\left|\sum\limits_x \varphi_x \right|\\
\end{array}
\right).
\eeq
Here we use this second approach. According to the notation in \eq{eq:ContinuumPhiAction}, which already 
includes a factor $1/2$, the relation between the vev $v$ and the expectation value of $\varphi^{rot}$
is then given as 
\beq
\label{eq:DefOfVEV}
\langle \varphi^{rot} \rangle = 
\left(
\begin{array}{*{1}{c}}
0\\
v\\
\end{array}
\right).
\eeq
In this setup the unrenormalized Higgs mode $h_x$ and the Goldstone modes $g^1_x,g^2_x,g^3_x,$ can then 
directly be read out of the rotated field according to
\beq
\label{eq:DefOfHiggsAndGoldstoneModes}
\varphi_x^{rot}  = 
\left(
\begin{array}{*{1}{c}}
g_x^2 + ig_x^1\\
v + h_x - i g_x^3\\
\end{array}
\right).
\eeq
The great advantage of this approach is that no limit procedure $J\rightarrow 0$ has to be performed, which
simplifies the numerical evaluation of the model tremendously. 

The so far lacking prescriptions for calculating the Goldstone renormalization constant $Z_G$, the Higgs boson mass 
$m_H$, and the quark masses $m_{t',b'}$ have been discussed in detail in \Ref{Gerhold:2010bh}. Here it is only stated that 
$Z_G$ is computed from the slope of the inverse Goldstone propagator, the functional form of which has been discussed at
one-loop order in \Ref{Gerhold:THESIS}. The latter propagator $\tilde G_G(p)$ is defined as 
\bea
\tilde G_G(p) &=& \frac{1}{3}\sum\limits_{\alpha=1}^3 \langle \tilde g^\alpha_p \tilde g^\alpha_{-p}\rangle, \\
\tilde g^\alpha_p &=& \frac{1}{\sqrt{L_s^3\cdot L_t}}\sum\limits_x e^{-ipx} g^\alpha_x
\eea
at the discrete lattice momenta $p_\mu=2\pi n_\mu/L_{s,t}$, $n_\mu = 0, \hdots, L_{s,t}-1$. Correspondingly,
the Higgs boson mass $m_H$ is derived from the Higgs propagator given as
\bea
\tilde G_H(p) &=& \langle \tilde h_p \tilde h_{-p}\rangle, \\
\tilde h_p &=& \frac{1}{\sqrt{L_s^3\cdot L_t}}\sum\limits_x e^{-ipx} h_x.
\eea
Finally, it is remarked that the lattice results on the quark masses presented in this paper have been computed
from the decay of the fermionic time correlation functions $C_f(\Delta t)$ at large Euclidean time separations 
$\Delta t$, where $f=t',b'$ denotes the quark flavour here. On the lattice the latter time correlation functions 
can be defined as
\bea
\label{eq:DefOfFermionTimeSliceCorr}
C_f(\Delta t) &=& \frac{1}{L_t\cdot L_s^6} \sum\limits_{t=0}^{L_t-1} \sum\limits_{\vec x, \vec y}
\Big\langle \,\RE\,\Tr\,\left(f_{L,t+\Delta t, \vec x}\cdot \bar f_{R,t,\vec y}\right) \Big\rangle,
\eea
where the left- and right-handed spinors are given through the projection operators according to
\bea
\label{eq:DefOfLeftHandedSpinors}
\left(
\begin{array}{*{1}{c}}
t'\\
b'\\
\end{array}
\right)_L
= \hat P_- 
\left(
\begin{array}{*{1}{c}}
t'\\
b'\\
\end{array}
\right)
 &\mbox{ and }&  (\bar t', \bar b')_R  = (\bar t', \bar b') P_-.
\eea
It is remarked that the given fermionic correlation function would be identical to zero due to the exact lattice 
chiral symmetry obeyed by the considered Higgs-Yukawa model, if one would not rotate the scalar field $\varphi$ 
according to \eq{eq:GaugeRotation} as discussed above. This rotation is implicitly assumed in the following. 
Furthermore, it is pointed out that the full {\textit{all-to-all}} correlator as defined in \eq{eq:DefOfFermionTimeSliceCorr} 
can be trivially computed. This all-to-all correlator yields very clean signals for the $t',b'$ quark mass 
determination.

\section{Numerical Results}
\label{sec:numericalresults}

For the eventual determination of the cutoff dependent Higgs boson mass bounds several series of Monte-Carlo calculations 
have been performed at different values of $\Lambda$ and on different lattice volumes as summarized in \tab{tab:SummaryOfParametersForUpperHiggsMassBoundRuns}.
In order to tame finite volume effects as well as cutoff effects to an acceptable level, we have demanded as 
a minimal requirement that all particle masses $\hat m=m_{H}, m_{t'}, m_{b'}$ in lattice units fulfill $\hat m < 0.5$ 
and $\hat m\cdot L_{s,t}>3.5$, at least on the largest investigated lattice volumes. Assuming the Higgs boson mass $m_H$ 
to be around $\GEV{500-750}$ this allows to reach cutoff scales between $\GEV{1500}$ and $\GEV{3500}$ on a \latticeX{24}{32}{.} 
However, despite this setting strong finite volume effects are nevertheless expected induced by the massless Goldstone modes. 
It is known that these finite size effects are proportional to $1/L_s^2$ at leading order~\cite{Hasenfratz:1989ux,Hasenfratz:1990fu,Gockeler:1991ty}. 
An infinite volume extrapolation of the lattice data is therefore mandatory. 

\includeTab{|ccccccc|}
{
$\kappa$ & $L_s$       & $L_t$ &  $N_c$ &  $m_0^2$             & $\lambda$      & $y_{t'}=y_{b'}$      \\
\hline
0.09442  & 12,16,20,24 &   32  &  1     &  2.5910               & 0              & 3.2122                \\
0.09485  & 12,16,20,24 &   32  &  1     &  2.5430               & 0              & 3.2049                \\
0.09545  & 12,16,20,24 &   32  &  1     &  2.4767               & 0              & 3.1949                \\
0.09560  & 12,16,20,24 &   32  &  1     &  2.4603               & 0              & 3.1923                \\
0.09605  & 12,16,20,24 &   32  &  1     &  2.4112               & 0              & 3.1849                \\ \hline
0.21300  & 12,16,20,24 &   32  &  1     &  $\infty$            & $\infty$        & 3.3707                \\
0.21500  & 12,16,20,24 &   32  &  1     &  $\infty$            & $\infty$        & 3.3550                \\
0.22200  & 12,16,20,24 &   32  &  1     &  $\infty$            & $\infty$        & 3.1816                \\
0.22320  & 12,16,20,24 &   32  &  1     &  $\infty$            & $\infty$        & 3.1730                \\
0.22560  & 12,16,20,24 &   32  &  1     &  $\infty$            & $\infty$        & 3.1561                \\
}
{tab:SummaryOfParametersForUpperHiggsMassBoundRuns}
{The model parameters underlying the lattice calculations performed in this study are presented. The setting $\lambda=0$ ($\lambda=\infty$) is employed
for deriving the lower (upper) Higgs boson mass bound. Depending on the lattice volume the available statistics ranges from $\Nconf=1,000$ to
$\Nconf=20,000$.
}
{Model parameters of the Monte-Carlo runs underlying the lattice calculation of the upper and lower Higgs boson mass bounds.}

The finite volume data of the renormalized vacuum expectation value $v_r$, the Higgs boson mass $m_H$, and the degenerate quark mass 
$m_{t'}=m_{b'}$ resulting from the calculations specified in \tab{tab:SummaryOfParametersForUpperHiggsMassBoundRuns}
are presented in \fig{fig:FiniteSizeEffects}. These lattice data are plotted versus $1/L_s^2$ and 
extrapolated to the infinite volume limit by means of a linear fit ansatz according to the aforementioned leading order behaviour. 
Due to the observed curvature arising from the non-leading finite volume corrections only those volumes with $L_s\ge 16$ have been 
respected by the linear fit procedures. One finds that the intended infinite volume extrapolation can indeed reliably be performed 
thanks to the multitude of investigated lattice volumes reaching from $12^3\times 32$ to \lattices{24}{32} here.

\includeFigTripleDouble{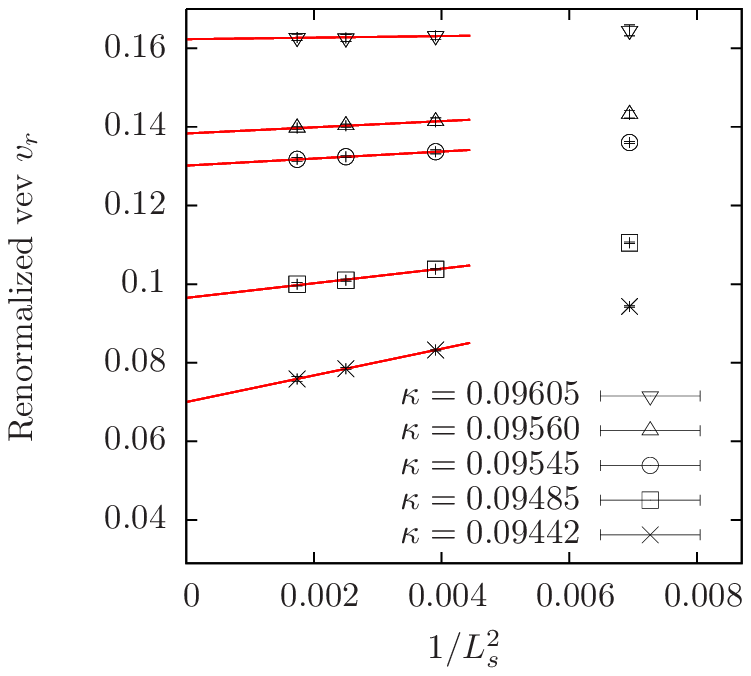}{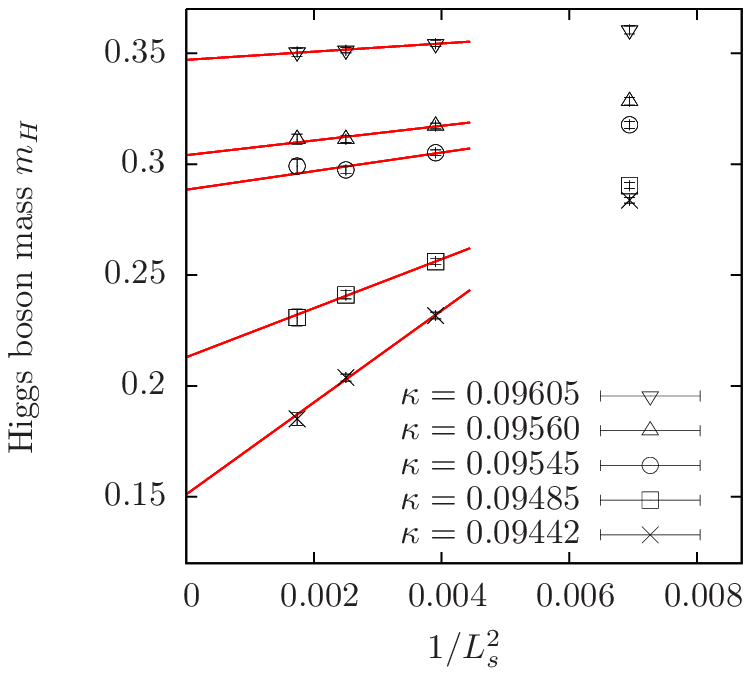}{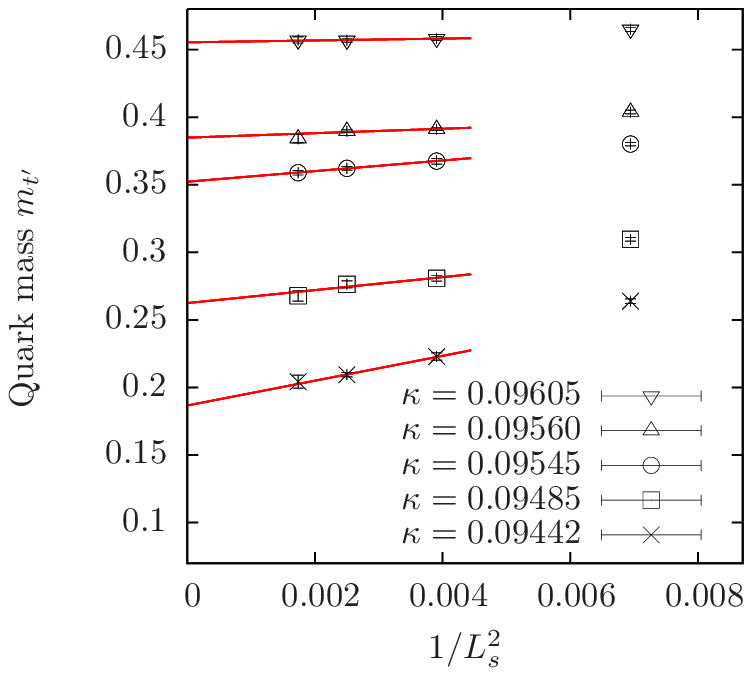}
{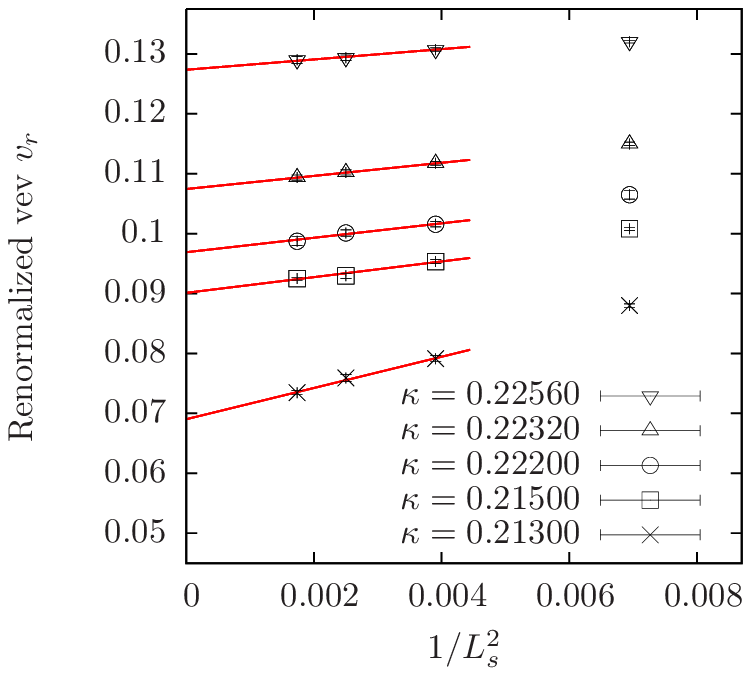}{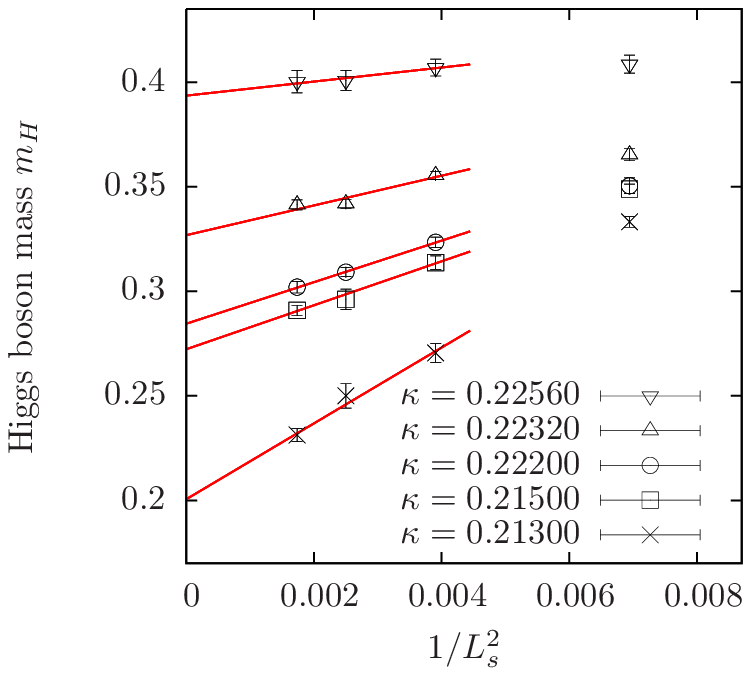}{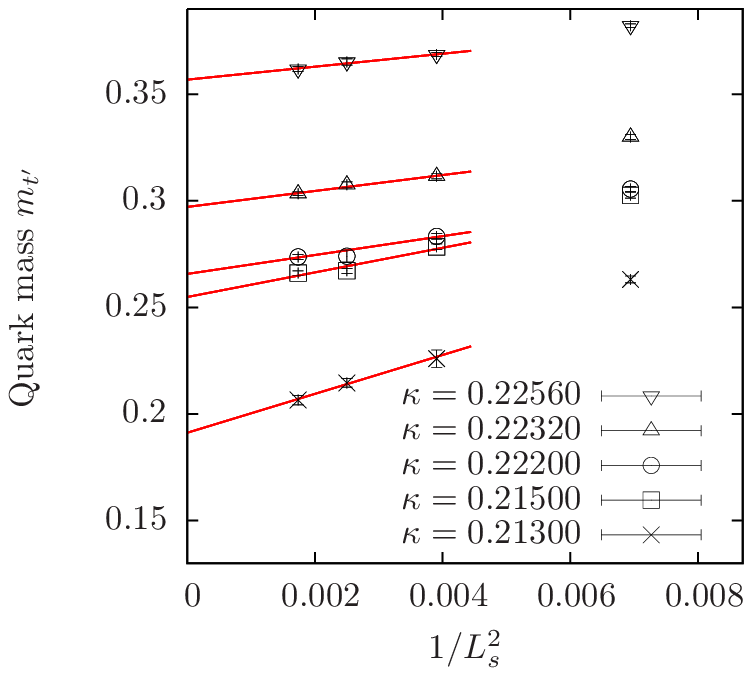}
{fig:FiniteSizeEffects}
{The finite volume data of the renormalized vacuum expectation value $v_r$ (a,d), the Higgs boson mass $m_H$ (b,e), and the degenerate quark mass 
$m_{t'}=m_{b'}$ (c,f), as obtained from the lattice calculations specified in \tab{tab:SummaryOfParametersForUpperHiggsMassBoundRuns},
are plotted versus $1/L_s^2$. The upper (lower) row corresponds to the setting $\lambda=0$ ($\lambda=\infty$). The infinite volume 
extrapolation is performed by fitting the data to a linear function. Due to the observed curvature arising from the non-leading finite
volume corrections only those volumes with $L_s\ge 16$ have been respected by the linear fit procedures.
}
{FiniteVolumeLatticeDataLowerBound}

The quality of the tuning procedure intending to hold the quark masses constant can then be examined in \fig{fig:PhysicalTopMasses}b displaying
the results of the infinite volume extrapolation of $m_{t'}$. In the considered SM4 scenario the fluctuation of the quark mass has been constrained 
to roughly $m_{t'}=m_{b'}=\GEV{676\pm 22}$. For later comparisons with the corresponding SM3 scenario we also present the analogous summary plot of 
our earlier investigations~\cite{Gerhold:2009ub,Gerhold:2010bh} of the latter setup where the degenerate quark masses have been fixed to approximately 
$m_t=m_b=\GEV{173\pm 3}$ as demonstrated in \fig{fig:PhysicalTopMasses}a.

\includeFigDouble{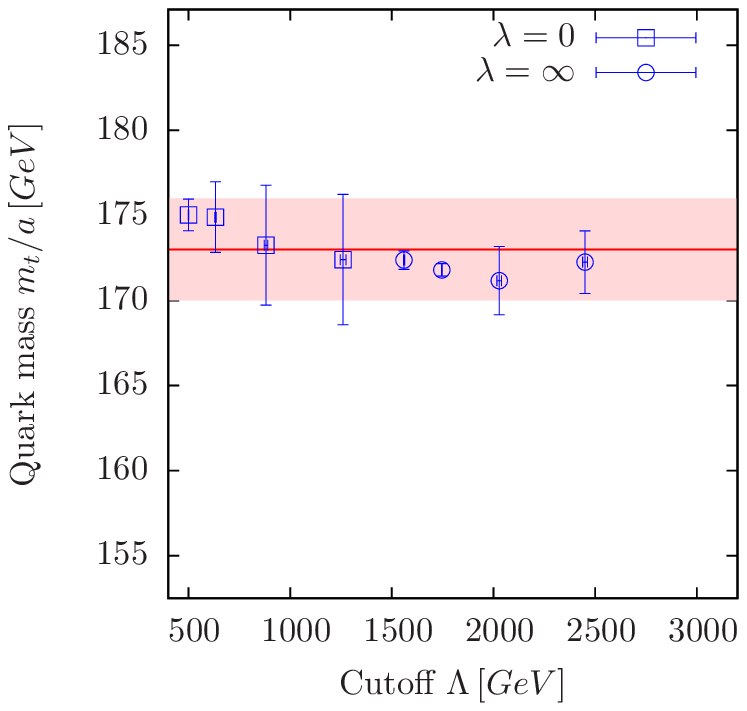}{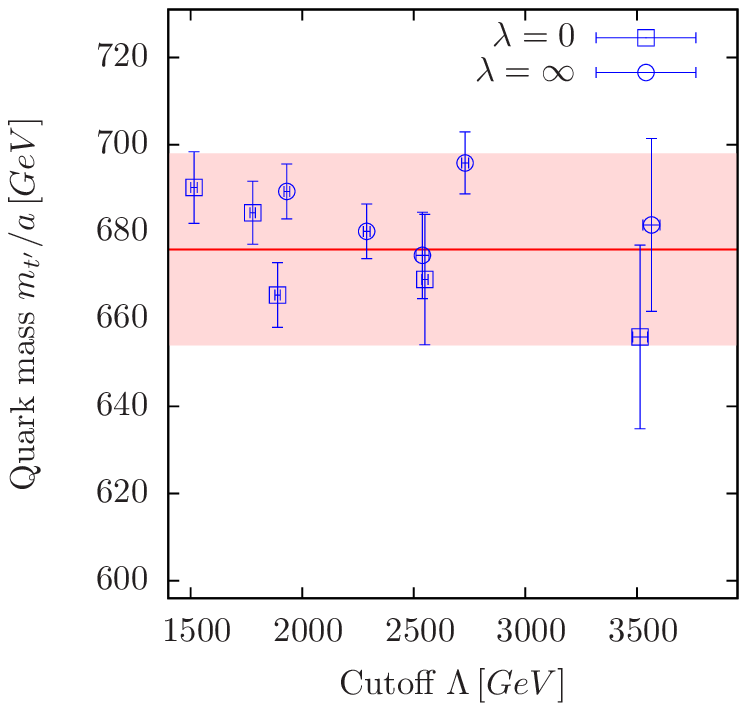}
{fig:PhysicalTopMasses}
{The infinite volume extrapolations of the degenerate quark masses observed in the lattice calculations specified in \tab{tab:SummaryOfParametersForUpperHiggsMassBoundRuns}
are presented versus the cutoff parameter $\Lambda$. In the SM3 scenario (a) the fluctuation of the quark mass has been constrained to $m_t=m_b=\GEV{173\pm 3}$, 
while $m_{t'}=m_{b'}=\GEV{676\pm 22}$ is adjusted in the SM4 scenario (b). 
}
{InfiniteVolumeExtrapolationBothBounds}{3}{1} 

The infinite volume results of the Higgs boson masses are finally presented in \fig{fig:PhysicalHiggsMassBounds}b. The numerical data for the
upper mass bound have moreover been fitted with the analytically expected functional form of the cutoff dependence of the upper
Higgs boson mass bound derived in \Ref{Luscher:1988uq}. It is given as
\bea
\label{eq:StrongCouplingLambdaScalingBeaviourMass}
\frac{m^{up}_{H}}{a} &=& A_m \cdot \left[\log(\Lambda^2/\mu^2) + B_m \right]^{-1/2}, 
\eea
with $A_m$, $B_m$ denoting the free fit parameters and $\mu$ being an arbitrary scale, which we have chosen as 
$\mu=\TEV{1}$ here. One learns from this presentation that the obtained results are indeed in good agreement 
with the expected logarithmic decline of the upper Higgs boson mass bound with increasing cutoff parameter $\Lambda$.

The reader may want to compare these findings to the upper and lower Higgs boson mass bounds previously derived in the SM3. The lattice 
results corresponding to that setup have been determined in \Ref{Gerhold:2009ub,Gerhold:2010bh} and are summarized in \fig{fig:PhysicalHiggsMassBounds}a.
The main finding is that especially the lower bound is drastically shifted towards larger values in the presence of the assumed mass-degenerate 
fourth quark doublet. The upper bound is also significantly increased by the fermionic contributions, however less strongly. From this analysis it 
seems conclusive that the usually expected light Higgs boson is incompatible with a very heavy fourth fermion generation.

\includeFigDouble{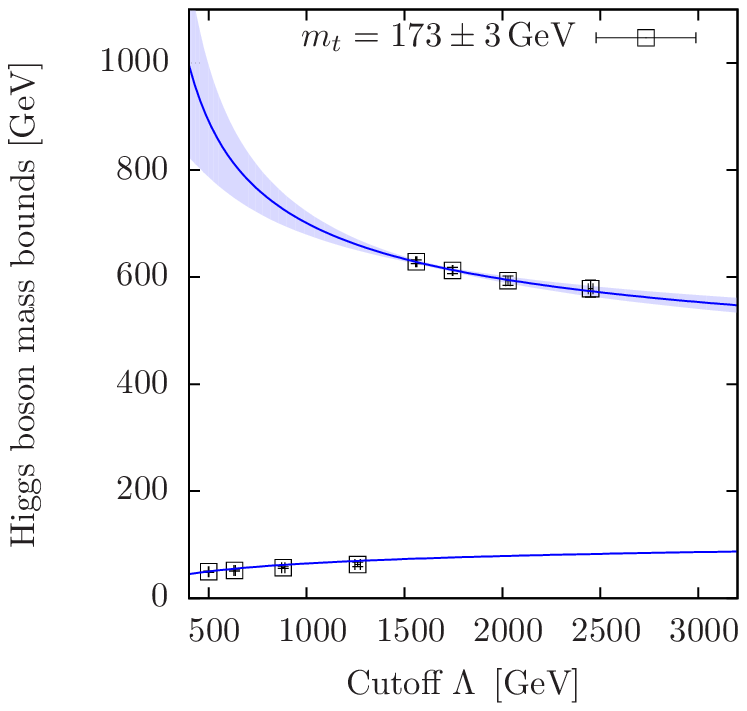}{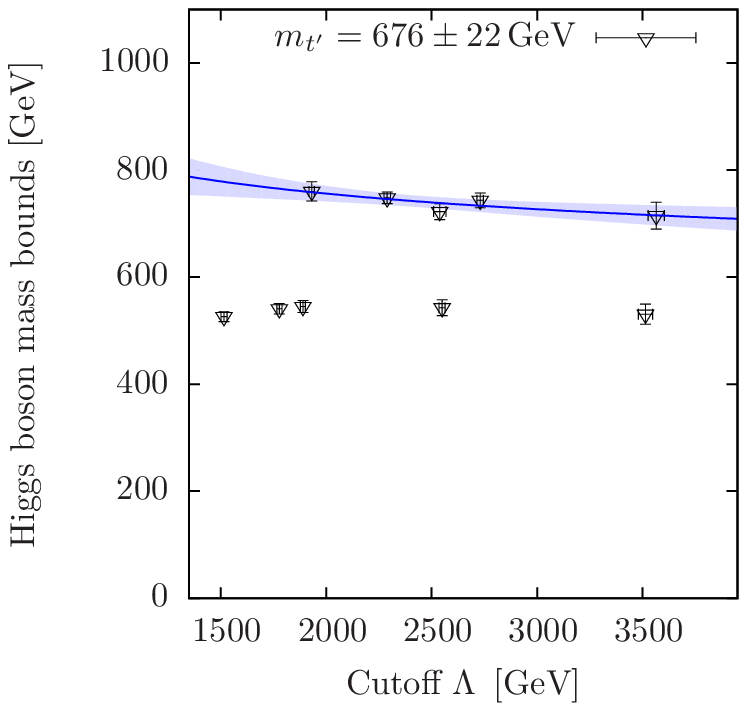}
{fig:PhysicalHiggsMassBounds}
{Upper and lower Higgs boson mass bounds are shown for $N_c=1$, $m_t=m_b=\GEV{173\pm 3}$ (a) and $N_c=1$, $m_{t'}=m_{b'}=\GEV{676\pm 22}$ (b).
Both upper bounds are each fitted with \eq{eq:StrongCouplingLambdaScalingBeaviourMass}. The lower bound in (a)
is also compared to a direct analytical computation depicted by the solid line as discussed in \Ref{Gerhold:2009ub}.
}
{InfiniteVolumeExtrapolationBothBounds}{3}{1}

\section{Outlook and Conclusions}
\label{sec:conclusions}

The aim of the present work has been the non-perturbative determination of the cutoff dependent upper and lower mass bounds 
of the Higgs boson assuming the potential existence of a very heavy fourth quark generation. Due to the potentially strong 
coupling nature of the associated Yukawa coupling constants the lattice approach has been employed here to allow for a 
non-perturbative investigation of a Higgs-Yukawa model. It serves here as a reasonable simplification of the extended Standard Model
SM4, including a fourth fermion generation. The idea is that the considered model contains only those fields and interactions 
which are most essential for the generation of the Higgs boson mass, \ie the scalar and quark fields and their mutual interactions.
The model has been built on the basis of L\"uscher's proposal~\cite{Luscher:1998pq} for the construction 
of chirally invariant lattice Higgs-Yukawa models adapted, however, to the situation of $\varphi$ being a complex doublet equivalent 
to one Higgs and three Goldstone modes. The resulting chirally invariant lattice Higgs-Yukawa model, based here on the Neuberger 
overlap operator, then obeys a global $\SUtwoTimesUoneY$ symmetry, as desired. 

The fundamental strategy underlying the determination of the cutoff dependent Higgs boson mass bounds has then been the numerical 
evaluation of the maximal interval of Higgs boson masses attainable within the considered Higgs-Yukawa model in consistency with the 
assumed physical scenario in terms of the masses of the fourth quark generation. Owing to the existence of a fluctuating complex phase 
in the non-degenerate case~\cite{Gerhold:THESIS}, the $t',b'$ quark masses have been assumed to be degenerate in this work. Here we 
have considered a hypothetical physical scenario with degenerate fourth generation quark masses at roughly \GEV{700}. To fix the physical 
scale of the performed lattice calculations we have exploited the phenomenologically known value of the renormalized vacuum expectation 
value (vev) of the scalar field 

The aforementioned procedure has actually been simplified in this study by exploiting the - perturbatively expected and numerically
confirmed~\cite{Gerhold:2010bh} - knowledge that the largest attainable Higgs boson masses are indeed observed in the case of an infinite bare 
quartic coupling constant, while the smallest masses are obtained in the limit of vanishing bare quartic coupling constant. Consequently, 
the search for the upper and lower Higgs boson mass bounds has been constrained to the bare parameter settings $\lambda=\infty$ and 
$\lambda=0$, respectively. The resulting finite volume lattice data on the Higgs boson mass, the vacuum expectation value of
the scalar field, and the quark masses turned out to be sufficiently precise to allow for their reliable infinite volume 
extrapolation. The arising extrapolation results are moreover sufficiently precise to resolve their cutoff dependence as 
demonstrated in \fig{fig:PhysicalHiggsMassBounds}. Concerning the upper bound the obtained cutoff dependence is in good 
agreement with the analytically expected logarithmic decline and thus with the triviality picture of the Higgs-Yukawa sector. 
The main finding, however, is that the upper and lower bounds are both shifted towards larger values. While the upper bound is only mildly 
shifted by roughly \proz{25}, the lower bound is shifted very drastically to roughly \GEV{500}. From this analysis it seems conclusive 
that the usually expected light Higgs boson is incompatible with a very heavy fourth fermion generation.

For the future several connected questions should be further investigated. Firstly, it would be interesting to repeat the present study 
at several other settings of the fourth generation quark masses to compute the quark mass dependence of the Higgs boson mass bounds
systematically. This should be done in particular for the lower bound. Moreover, the implemented model allows to tune the bare coupling 
parameters to arbitrarily large values. It would be of particular interest to determine the largest possible quark mass attainable within 
the Higgs-Yukawa model beyond a perturbative unitarity consideration. This can be done by evaluating the model in the limit of infinite 
bare Yukawa coupling constants.

\section*{Acknowledgments}
We thank George Hou and David Lin for ongoing discussions and M. M\"uller-Preussker for his continuous support.
We moreover acknowledge the support of the DFG through the DFG-project {\it Mu932/4-2}.
The numerical computations have been performed on the {\it HP XC4000 System}
at the {Scientific Supercomputing Center Karlsruhe} and on the
{\it SGI system HLRN-II} at the {HLRN Supercomputing Service Berlin-Hannover}.

\bibliographystyle{unsrtOWN}
\bibliography{HiggsBosonMassBoundsAtMtop700GEV}

\end{document}